\documentclass[english,aps,secnumarabic,amsmath,amssymb,twocolumn,showpacs]{revtex4}
\usepackage[T1]{fontenc}
\usepackage[latin9]{inputenc}
\usepackage[active]{srcltx}
\usepackage{textcomp}
\usepackage{amssymb}
\usepackage{graphicx}

\makeatletter
\@ifundefined{textcolor}{}
{%
 \definecolor{BLACK}{gray}{0}
 \definecolor{WHITE}{gray}{1}
 \definecolor{RED}{rgb}{1,0,0}
 \definecolor{GREEN}{rgb}{0,1,0}
 \definecolor{BLUE}{rgb}{0,0,1}
 \definecolor{CYAN}{cmyk}{1,0,0,0}
 \definecolor{MAGENTA}{cmyk}{0,1,0,0}
 \definecolor{YELLOW}{cmyk}{0,0,1,0}
}



\usepackage{babel}
\begin{document}

\title{Engineering three-body interaction and Pfaffian states in circuit
QED systems}

\author{M. Hafezi} \email{hafezi@umd.edu}
\author{ P. Adhikari}
\author{J. M. Taylor}
\affiliation{Joint Quantum Institute, NIST/University of Maryland, College Park
MD}
\pacs{73.43.-f,85.25.Cp,05.30.Pr}
\begin{abstract}
We demonstrate a scheme to engineer the three-body interaction in circuit-QED
systems by tuning a fluxonium qubit. Connecting such qubits in a square
lattice and controlling the tunneling dynamics, in the form of a
synthesized magnetic field,  for the photon-like excitations of the system, allows the implementation of a parent
Hamiltonian whose ground state is  the Pfaffian wave function. Furthermore,
we show that the addition of the next-nearest neighbor tunneling stabilizes
the ground state, recovering the expected topological degeneracy
even for small lattices. Finally, we discuss the implementation of
these ideas with the current technology. 
\end{abstract}
\maketitle
Many-body topological states have fascinating properties such as non-Abelian
statistics that have been theoretically predicted but have not been
observed \cite{Nayak:2008p41368}. Such states have been also proposed
as a promising platform to perform robust quantum computation \cite{kitaev03}.
The simplest state with non-abelian properties was constructed by Moore and Read in the form of a Pfaffian wavefunction, in the context of fractional
quantum Hall effect \cite{MOORE:1991p15997}. As the same time, ``parent Hamiltonians''
have been introduced to generate such states as their ground states.
In particular, a Hamiltonian with three-body interaction was proposed
by Greiter et al. which yields the Pfaffian state \cite{Greiter:1991}. There
have been remarkable efforts to generate such Hamiltonians, e.g., using ultra
cold atom system \cite{Bloch:Review,Cooper:2008p6565}, however, the elimination 
 of the two-body interaction while preserving the bosonic nature of excitations remains challenging  \cite{Buchler:2007dk,Paredes:2007bi,Daley:2009p46211,Mazza:2010p37515,Mahmud:2013wf}, as expected for perturbatively-generated three-body terms \cite{Jordan:2008}.
In this Letter, we present a novel scheme using circuit QED systems  with ultra-strong microwave nonlinearity \cite{Schoelkopf:2008p8712,Stajic:2013dh}
to achieve this end. We demonstrate how to engineer a three-body interaction
and a synthetic magnetic field required to implement the parent
Hamiltonian of Ref.~\cite{Greiter:1991} in a lattice.

\begin{figure}
\includegraphics[width=0.4\textwidth]{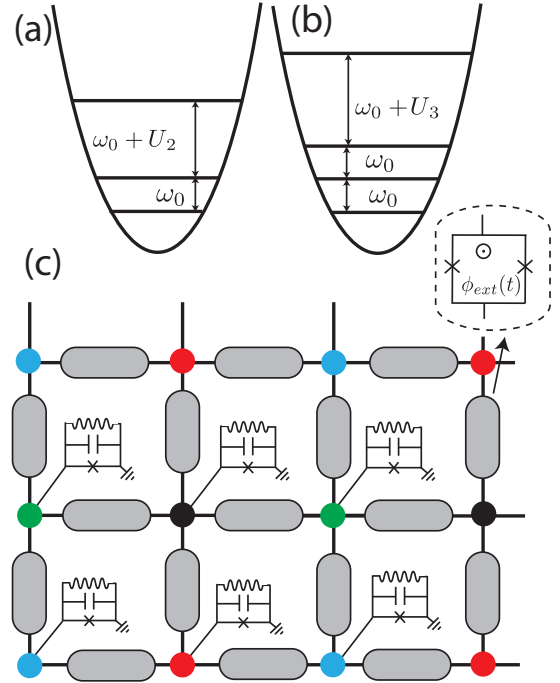}

\caption{(a) Two-body and (b) three-body interaction emerge as two- and three-excitation
nonlinearity in an anharmonic oscillator. (c) Circuit model to implement
Eq.(1). Each site is composed of an anharmonic oscillator with the three-body
interaction. The qubits are detuned from each other according to a
staggered pattern (blue/red/green/black) and are coupled by externally modulated
SQUIDs.}
\end{figure}

The key idea in this Letter is to introduce a generalized qubit that
exhibits the three-body interaction, i.e. one and two excitations are
allowed in the qubit but the creation of the third excitation has
an energy penalty, as shown in Fig.~1(b), compared to Fig.~1(a) for two-body interaction. Such qubit can be characterized
by a Hamiltonian of the form $H_{i}=U_{3}\hat{a}_{i}^{\dagger3}\hat{a_{i}}^{3}$,
where $\hat{a}_{i}$ is the bosonic creation operator at site
$i$ and $U_{3}$ is the interaction strength. This qubit can be generated
by tuning various parameters of a fluxonium qubit \cite{Manucharyan:2009fo},
in a parameter regime similar to that of a transmon \cite{Transom:2007},
to achieve the desired level structure. We couple these qubits in a square lattice, using SQUIDs where their external phase
is modulated, as shown in Fig.~1(c). This modulation can imprint a tunneling
phase \cite{ZakkaBajjani:2011in}, which is arranged to implement a synthetic magnetic field with
a fixed gauge. Such a system can be described by the following Hamiltonian
\begin{eqnarray}
H & = & -J\sum_{x,y}\hat{a}_{x+1,y}^{\dagger}\hat{a}_{x,y}e^{-i2\pi\alpha y}+\hat{a}_{x,y}^{\dagger}\hat{a}_{x+1,y}e^{+i2\pi\alpha y}\nonumber \\
 & + & \hat{a}_{x,y+1}^{\dagger}\hat{a}_{x,y}+\hat{a}_{x,y}^{\dagger}\hat{a}_{x,y+1}+\frac{1}{6}U_{3}\hat{a}_{x,y}^{\dagger3}\hat{a}_{x,y}^{3},\label{eq:parent_hamiltonian}
\end{eqnarray}
where $\alpha$ is the acquired phase from tunneling around a single
plaquette and characterizes the strength of the magnetic field. In
the continuum limit ($\alpha\ll1$), this Hamiltonian is known to
be the parent Hamiltonian of Pfaffian states, when the filling factor
is $\nu=N/N_{\alpha}=1$, where $N_{\alpha}$ is the number of particles
inside the system and $N_{\alpha}$ is the total number of magnetic
flux. In other words, the Pfaffian state is the ground state and the
excited states have non-abelian statistics. We present our numerical
results that indicate that indeed the ground state is the Pfaffian
state with the three-fold topological degeneracy. We show that adding
next-nearest-neighbor tunneling can flatten the single-particle energy
bands, and therefore, the Pfaffian state can be seen even for high
magnetic fields $(\alpha=0.25)$.

In order to implement the Hamiltonian of Eq.(1), we need two key
elements: (1) inducing a magnetic-type hopping between sites and (2)
generating the three-body nonlinearity. We start by describing a single
qubit that exhibits the three-body interaction and return to the discussion
of the magnetic hopping later. We consider a fluxonium \cite{Manucharyan:2009fo},
which can be described by the following Hamiltonian:

\begin{equation}
H_{qubit}=4 E_{c}n^{2}+\frac{1}{2}E_{L}\phi^{2}-E_{J}\cos(\phi+\phi_{x}),
\end{equation}
where $E_{c}$ is the single-electron charging energy, $E_{J}$ is the Josephson junction energy, and $E_{L}=(\Phi_0/2\pi)^2/L$ characterizes the
shunted inductive energy, defined in terms of the flux quantum $\Phi_0=h/2e$. $(n,\phi)$
are conjugate variables and are equal to $^{4}\sqrt{8 E_{c}/E_{L}}$(Cooper pair number,
node flux), respectively. $\phi_{x}$ is the external flux through
the Josephson junction in unit of the magnetic flux quantum. The spectrum of this Hamiltonian, which
basically describes a particle in a potential, can be numerically
obtained. We consider the so-called transmon regime where $E_{J}\gg E_{c}$
\cite{Transom:2007}, so that the qubit could remain less sensitive to charge
noise. We are interested in the limit where the first and the second
excitation levels have the same energy and the third excitation is
detuned from them (Fig.~1(b)). The nonlinearity is provided by the
Josephson junction and tuned to the desired form using the shunted 
inductor and the external flux. In particular, we analyze the four
lowest energy eigenstates. In such subspace, one can describe the
system with a general Hamiltonian of the form: 

\begin{equation}
H_{model}=\omega_{0}\hat{a}^{\dagger}\hat{a}+\frac{1}{2}U_{2}\hat{a}^{\dagger2}\hat{a}^{2}+\frac{1}{6}U_{3}\hat{a}^{\dagger3}\hat{a}^{3}
\end{equation}
where $\hat{a}^\dagger$ is the creation operator of a single excitation, $\omega_{0}$ is the energy of the lowest level and $U_{2}(U_{3})$ characterizes the two- (three-) particle interaction, as shown in
Fig.~1(a,b). 

Fig.~2 (a,b) shows the numerical results for $(U_{2},U_{3}),$ respectively.
We observe that for a given $E_{L}$, the external flux $\phi_{x}$
can be tuned so that the two-body nonlinearity vanishes. However,
the three-body nonlinearity does not necessarily vanishes for that
specific $\phi_{x}.$ Fig.~2(c) shows the value of the three-body nonlinearity
when we operate at $U_{2}=0$. The largest value of $U_3$, while keeping the bosonic nature of the excitations (see the next paragraph), is achieved for the following parameters: $E_{L}\simeq1.4 E_{J},\phi_{x}\simeq2.68$, for $E_{c}=0.05 E_{J}$. This suggests that once such qubits are coupled to each other, the
excitations can hop in between them, and only zero, one and two Fock
states on each site can be occupied. However, it is not guaranteed
that the hopping has the correct bosonic form. 

\begin{figure}
\includegraphics[width=0.4\textwidth]{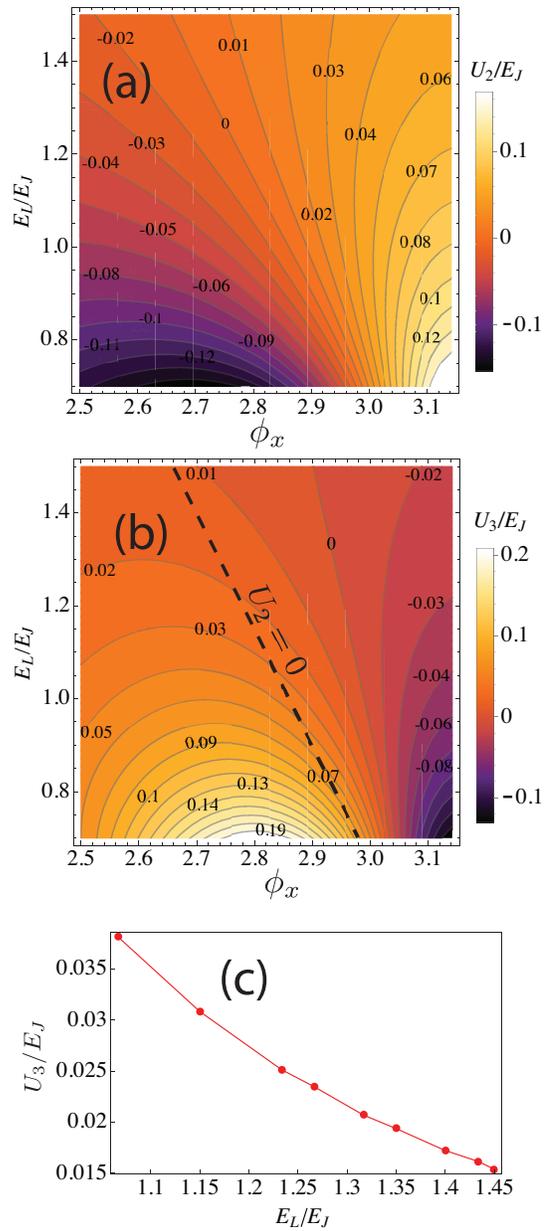}\caption{(a) Three-body ($U_3$) and (b) two-body ($U_2$) interaction strength as function of the
external bias flux and the shunted inductance. The dashed line on (b) shows where the two-body interaction vanishes. (c) shows the three-body
interaction strength when $U_{2}$ is optimized to be less than $0.0005 E_{J}$. All the plots are for $E_{c}/E_{J}=0.05.$}
\end{figure}

In order to verify that the hopping has the correct bosonic enhancement
factor (for example $ \hat{a}|n\rangle=\sqrt{n}|n-1\rangle$), we consider two qubits that are inductively coupled to each
other (Fig.~3(a)), and analyze the dynamics of several excitations in between them.
In particular, we consider the coupling of the form: $H_{coup}=M \phi_{1}\phi_{2}$,
where $M$ is the tunneling energy, proportional to the mutual inductance,
and $\phi_{i}$ is the phase of the qubit $i$. First, we study the
dynamics of a single excitation in the coupled system. We prepare
one qubit in a Fock state with one excitation and let the system evolve.
The population dynamics is plotted in Fig.~4(b). We observe that the
system undergoes Rabi oscillations between two states. Next, we consider
two excitations in the system. Similarly, we initiate the system with
the two-excitation Fock state and study the dynamics. We see that
the microscopic Hamiltonian of our system leads to a dynamic identical
to that of a model system of two bosonic oscillators with two excitations, as shown in Fig.~3(c).
Finally, we start with three excitations in the system, one in the
first qubit and two in the second qubit, as shown in Fig.~3(d). Due to the presence of the three-body nonlinearity ($U_{3}\neq0$$),$ we observe the population in the
three-excitation Fock states is suppressed (less than $10^{-7}$),
similar to a model system of two bosonic systems in the three-body
hard-core limit ($\hat{a}_{i}^{3}|\Psi\rangle=0$, where $|\Psi\rangle$
is an arbitrary state of the system). 
\begin{figure}
\includegraphics[width=0.48\textwidth]{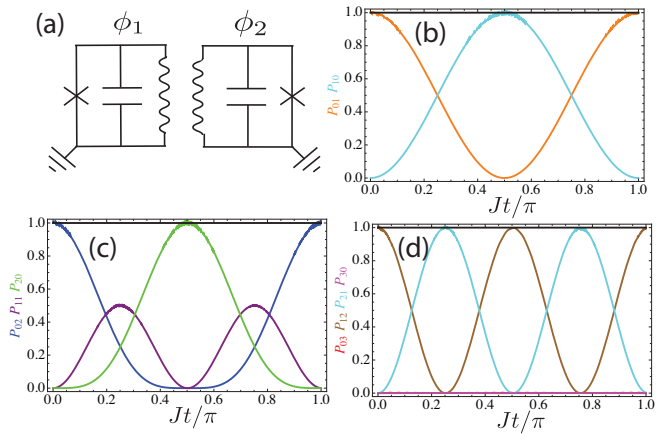}\caption{Dynamics of different excitation numbers when two qubits are inductively
coupled to each other, as shown in (a): (b) one excitation, (c) two
excitations (d) three excitations. The black line shows the total population in (b) and (c) and $P_{21}+P_{12}$ in (d).}
 
\end{figure}

Now, we discuss the implementation of the magnetic hopping terms in
Eq.(1). There have been several proposals in the past to engineer magnetic-like
Hamiltonian in the context of circuit-QED systems \cite{Koch:2010,Kapit:2013wx} and 
also proposals without breaking the time reversal symmetry in photonic systems \cite{Hafezi:2011delay,Umucalilar:2011}.
Here, we present a scheme based on phase modulating the SQUIDs that
couple adjacent sites. In particular, we set
the resonance ($\omega_i$) of adjacent sites to be different than each other and form a staggered pattern, as shown in Fig.~1(c). The connecting SQUID inductance
is modulated by applying a microwave flux $\phi_{ext}(t)=\delta\phi\cos(\Delta_{ij} t+\phi_{p})$,
in unit of the magnetic  flux quantum, where $\delta\phi\ll1$
and the microwave pump frequency is tuned to the frequency difference
of two adjacent sites ($\Delta_{ij}=\omega_i-\omega_j$). As shown in Ref. \cite{ZakkaBajjani:2011in},
such modulation induces a hopping Hamiltonian between two modes of
the form $a_{i}^{\dagger}a_{j}e^{+i\phi_{p}}+a_{j}^{\dagger}a_{i}e^{-i\phi_{p}}$, in the rotating frame with the rotating wave approximation.
The difference between our case and Ref.\cite{ZakkaBajjani:2011in}
is that, there the hopping was induced between two modes of the same
waveguide, while here the hopping is induced between two modes of different
sites. In contrast to the previous scheme proposed by Koch et al. \cite{Koch:2010}, our scheme is not sensitive to charge noise, and the generated magnetic field is insensitive to minor device variations.

The Pfaffian state is the ground state of the Hamiltonian of Eq.(1) in
the continuum limit ($\alpha\ll1$). However, the lattice could distort
the wave function, close the gap and destroy the topological order. To map the
lattice to the continuum in such models, the concept of long-range tunneling
was suggested by Laughlin \cite{Laughlin:1989kd}. By introducing the long-range hopping,
the single particle spectrum becomes flat and the many-body gap is
enhanced. As an example, for Laughlin fraction ($\nu=1/2$) in bosonic
systems, for large magnetic field $(\alpha\gtrsim.4) $\cite{Hafezi:PRA2007},
the gap closes and the topological order of the ground state disappears.
However, Kapit and Mueller showed that including long-range order
tunneling flattens the lowest branch of the single particle Hofstadter's
spectrum and improves the gap even for large magnetic fields $\alpha\simeq0.5$ \cite{Kapit:2010p47218}.
Furthermore, the long range tunneling allows braiding operation even
for small lattices \cite{Kapit:2012bn}. Here, we consider the
Pfaffian fraction for bosons ($\nu=1$) and we observe similar behavior. We assume the bosonic occupation number on each site
can not exceed two, i.e., $U_{3}=\infty$. In this situation, the
ground state of the system on the torus should be three-fold degenerate
and should be separated by a gap \cite{Greiter:1991}. As shown in
Fig.~4(a), and previously reported in Ref. \cite{Mazza:2010p37515},
the gap is non-zero even in a small 4x4 lattice. The ground state
has the expected order, i.e., the Chern number is equal to one for
the ground state manifold and a very weak overlap with the Pfaffian
state, as expected for the lattice. Now, we consider the next-nearest-neighbor
tunneling terms which in the Landau gauge are described as, 

\begin{eqnarray*}
\sum_{x,y}\sum_{\Delta x,\Delta y} &  & (-1)^{\Delta x+\Delta y+\Delta x\Delta y}e^{-\frac{\pi}{2}(1-\alpha)(\Delta x^{2}+\Delta y^{2})}\\
 & \times & e^{-i2\pi\alpha(y\Delta x+\Delta x\Delta y/2)}\hat{a}_{x+\Delta x,y+\Delta y}^{\dagger}\hat{a}_{x,y}
\end{eqnarray*}
where $\Delta x(\Delta y)$ is the number of site tunneling in the
x(y) direction, respectively. If we consider only $(\Delta x=1,\Delta y=0)$
and $(\Delta x=0,\Delta y=1)$ terms, we recover the tunneling terms
in Eq.(1). Since our implementation of magnetic field with modulation requires two connecting sites to have different frequencies, we choose a staggered patterned with four colors, as shown in Fig.~1(c), to implement the next-nearest-neighbor tunneling terms. As shown in Fig.~4(a), if we include next-nearest-neighbor terms, the three-fold ground state degeneracy is preserved and the gap is improved three folds. In contrast
to atoms on optical lattices, here in circuit-QED systems, the long
range tunneling term can be implemented by linking different sites
using extra connecting SQUIDs.

In an experimental realization, one might not be able to access very
large three-body interaction $(U_{3}\gg J$), and entirely suppress
the two-body interaction $(U_{2}=0)$, therefore, it is important
to assure that the gap between the three-fold degenerate ground state manifold
and the excited state is preserved for a non-ideal situation.
To numerically investigate that, we define an ``order parameter'',
as the ratio between the gap and the energy difference within the
ground state, i.e. $\lambda=(E_{4}-E_{3})/(E_{3}-E_{1})$, where $E_i$ is the energy of the $i$-th eigenstate. As we see
in Fig.~4(b), the system is completely gapped for large $U_{3}$ and
zero $U_{2}$. As $U_{3}$ becomes small and $U_{2}$ becomes large,
the order parameter vanishes, the gap closes. Therefore, we see that
for a certain region in the parameter space the gap exists and the
system is robust against the presence of a finite two-body interaction. 

\begin{figure}

\includegraphics[scale=0.4]{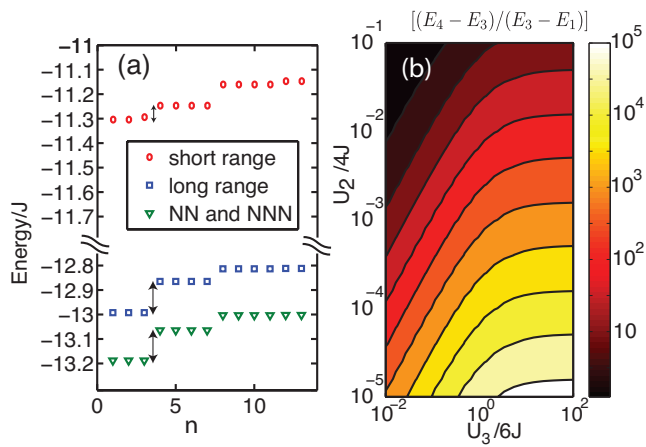}\caption{(a) First thirteen eigenvalues for 4 particles on a 4\texttimes{}4
lattice with torus boundary conditions, $\alpha=0.25$, and $U_{3}=\infty$.
The index n labels eigenvalues. Different shapes show different tunneling
schemes: Circles ($\bigcirc$) for only nearest neighbor tunneling
(gap/J$\simeq$0.04), triangles ($\bigtriangledown)$ for the nearest
and the next-nearest-neighbor tunneling (gap/J$\simeq$0.12) and squares
$(\square)$ for the long-range tunneling (gap/J$\simeq$0.12). The gaps are denoted by arrows. (b) The order parameter as a function of $U_3$ and $U_2$, characterizing the relative size of the gap.}
\end{figure}

We next consider experimental issues involving the realization and
detection of Pfaffian states in the proposed circuit QED system. As
we discussed above the three-body interaction should be larger than
the tunneling $U_{3}\gg J$, i.e., we should choose the largest (smallest)
possible $U{}_{3}$ ($J$), respectively. On the one hand, the three-body
interaction is bounded from above as a small fraction of the Josephson
energy. Considering a Josephson energy of tens of GHz, one can
achieve a three-body interaction strength of a few 100MHz. On the
other hand, $J$ is bounded from below, since the tunneling process
should be faster than any decoherence mechanism. In particular, $J\gg T_{2}^{-1},T_{1}^{-1}$
where $T_{1}(T_{2})$ is the relaxation (decoherence) time, respectively.
Recent experiments have shown $T_{1},T_{2}\gg10\mu s$ \cite{Stajic:2013dh},
therefore, assuming a tunneling rate of $J\geq (2\pi)10$ MHz guarantees
many tunnelings to occurs before the coherence is lost. Therefore,
one can achieve a regime where $U_{3}\gtrsim60J$ which according
to Fig.~4(b) provides the sufficient energy gap between the ground
manifold and the excited states. 

In summary, we have presented a scheme to implement the three-body interaction
and the Hamiltonian to generate the Pfaffian state in circuit-QED systems.
Due to relatively slow decoherence in these systems, one can
use an adiabatic approach to prepare such states \cite{Umucalilar:2011b,Hayward:2012,Hafezi:2013wh}
and locally probe each site using an auxiliary transmon \cite{Shanks:2013hi}.
However, a more relevant regime for such photonic systems is to externally
drive them and investigate their many-body non-equilibrium behavior \cite{Carusotto:2009,Hafezi:2012ku}
and analyze their incompressibility \cite{Nunnenkamp:2011dm,schiro:2012,Hafezi:2013wh,Umucalilar:2012vw}.
Another intriguing direction is to explore the braiding of non-abelian
anyons in such systems. It has been recently shown that even in small
systems with long-range tunneling such braiding can be obtained \cite{Kapit:2012bn}.
One can possibly dynamically detune the resonance frequency of the
site to trap and move around the anyons. 

This research was supported by the U.S. Army Research office MURI
award W911NF0910406 and NSF through the Physics Frontier Center at
the Joint Quantum Institute. We thank
L. Mazza, E. Kapit, A. Houck, A. Rimberg, F. Wellstood and B. Palmer for fruitful discussions.

We recently became aware of a similar proposal, concurrently developed by Kapit and Simon, using spin-1/2 systems (arXiv:1307.3485).

\bibliographystyle{apsrev}

\begin{thebibliography}{31}
\expandafter\ifx\csname natexlab\endcsname\relax\def\natexlab#1{#1}\fi
\expandafter\ifx\csname bibnamefont\endcsname\relax
  \def\bibnamefont#1{#1}\fi
\expandafter\ifx\csname bibfnamefont\endcsname\relax
  \def\bibfnamefont#1{#1}\fi
\expandafter\ifx\csname citenamefont\endcsname\relax
  \def\citenamefont#1{#1}\fi
\expandafter\ifx\csname url\endcsname\relax
  \def\url#1{\texttt{#1}}\fi
\expandafter\ifx\csname urlprefix\endcsname\relax\def\urlprefix{URL }\fi
\providecommand{\bibinfo}[2]{#2}
\providecommand{\eprint}[2][]{\url{#2}}

\bibitem[{\citenamefont{Nayak et~al.}(2008)\citenamefont{Nayak, Simon, Stern,
  Freedman, and Sarma}}]{Nayak:2008p41368}
\bibinfo{author}{\bibfnamefont{C.}~\bibnamefont{Nayak}},
  \bibinfo{author}{\bibfnamefont{S.}~\bibnamefont{Simon}},
  \bibinfo{author}{\bibfnamefont{A.}~\bibnamefont{Stern}},
  \bibinfo{author}{\bibfnamefont{M.}~\bibnamefont{Freedman}}, \bibnamefont{and}
  \bibinfo{author}{\bibfnamefont{S.}~\bibnamefont{Sarma}},
  \bibinfo{journal}{Rev. Mod. Phys.} \textbf{\bibinfo{volume}{80}},
  \bibinfo{pages}{1083} (\bibinfo{year}{2008}).

\bibitem[{\citenamefont{Kitaev}(2003)}]{kitaev03}
\bibinfo{author}{\bibfnamefont{A.}~\bibnamefont{Kitaev}},
  \bibinfo{journal}{Annals of Physics} \textbf{\bibinfo{volume}{303}},
  \bibinfo{pages}{2} (\bibinfo{year}{2003}).

  
\bibitem[]{MOORE:1991p15997}
\bibinfo{author}{\bibfnamefont{G.}~\bibnamefont{Moore}},
\bibnamefont{and}
  \bibinfo{author}{\bibfnamefont{N.}~\bibnamefont{Read}},
    \bibinfo{journal}{Nuc. Phys. B} \textbf{\bibinfo{volume}{360}},
  \bibinfo{pages}{362} (\bibinfo{year}{191}).  
  

\bibitem[{\citenamefont{Greiter et~al.}(1991)\citenamefont{Greiter, Wen, and
  Wilczek}}]{Greiter:1991}
\bibinfo{author}{\bibfnamefont{M.}~\bibnamefont{Greiter}},
  \bibinfo{author}{\bibfnamefont{X.-G.} \bibnamefont{Wen}}, \bibnamefont{and}
  \bibinfo{author}{\bibfnamefont{F.}~\bibnamefont{Wilczek}},
  \bibinfo{journal}{Phys. Rev. Lett.} \textbf{\bibinfo{volume}{66}},
  \bibinfo{pages}{3205} (\bibinfo{year}{1991}).

\bibitem[{\citenamefont{Bloch et~al.}(2008)\citenamefont{Bloch, Dalibard, and
  Zwerger}}]{Bloch:Review}
\bibinfo{author}{\bibfnamefont{I.}~\bibnamefont{Bloch}},
  \bibinfo{author}{\bibfnamefont{J.}~\bibnamefont{Dalibard}}, \bibnamefont{and}
  \bibinfo{author}{\bibfnamefont{W.}~\bibnamefont{Zwerger}},
  \bibinfo{journal}{Rev. Mod. Phys.} \textbf{\bibinfo{volume}{80}},
  \bibinfo{pages}{885} (\bibinfo{year}{2008}).

\bibitem[{\citenamefont{Cooper}(2008)}]{Cooper:2008p6565}
\bibinfo{author}{\bibfnamefont{N.}~\bibnamefont{Cooper}},
  \bibinfo{journal}{Advances in Physics} \textbf{\bibinfo{volume}{57}},
  \bibinfo{pages}{539} (\bibinfo{year}{2008}).

\bibitem[{\citenamefont{B{\"u}chler et~al.}(2007)\citenamefont{B{\"u}chler,
  Micheli, and Zoller}}]{Buchler:2007dk}
\bibinfo{author}{\bibfnamefont{H.~P.} \bibnamefont{B{\"u}chler}},
  \bibinfo{author}{\bibfnamefont{A.}~\bibnamefont{Micheli}}, \bibnamefont{and}
  \bibinfo{author}{\bibfnamefont{P.}~\bibnamefont{Zoller}},
  \bibinfo{journal}{Nat. Phys.} \textbf{\bibinfo{volume}{3}},
  \bibinfo{pages}{726} (\bibinfo{year}{2007}).

\bibitem[{\citenamefont{Paredes et~al.}(2007)\citenamefont{Paredes, Keilmann,
  and Cirac}}]{Paredes:2007bi}
\bibinfo{author}{\bibfnamefont{B.}~\bibnamefont{Paredes}},
  \bibinfo{author}{\bibfnamefont{T.}~\bibnamefont{Keilmann}}, \bibnamefont{and}
  \bibinfo{author}{\bibfnamefont{J.}~\bibnamefont{Cirac}},
  \bibinfo{journal}{Phys. Rev. A} \textbf{\bibinfo{volume}{75}},
  \bibinfo{pages}{053611} (\bibinfo{year}{2007}).

\bibitem[{\citenamefont{Daley et~al.}(2009)\citenamefont{Daley, Taylor, Diehl,
  Baranov, and Zoller}}]{Daley:2009p46211}
\bibinfo{author}{\bibfnamefont{A.~J.} \bibnamefont{Daley}},
  \bibinfo{author}{\bibfnamefont{J.~M.} \bibnamefont{Taylor}},
  \bibinfo{author}{\bibfnamefont{S.}~\bibnamefont{Diehl}},
  \bibinfo{author}{\bibfnamefont{M.}~\bibnamefont{Baranov}}, \bibnamefont{and}
  \bibinfo{author}{\bibfnamefont{P.}~\bibnamefont{Zoller}},
  \bibinfo{journal}{Phys. Rev. Lett.} \textbf{\bibinfo{volume}{102}},
  \bibinfo{pages}{040402} (\bibinfo{year}{2009}).

\bibitem[{\citenamefont{Mazza et~al.}(2010)\citenamefont{Mazza, Rizzi,
  Lewenstein, and Cirac}}]{Mazza:2010p37515}
\bibinfo{author}{\bibfnamefont{L.}~\bibnamefont{Mazza}},
  \bibinfo{author}{\bibfnamefont{M.}~\bibnamefont{Rizzi}},
  \bibinfo{author}{\bibfnamefont{M.}~\bibnamefont{Lewenstein}},
  \bibnamefont{and} \bibinfo{author}{\bibfnamefont{J.~I.} \bibnamefont{Cirac}},
  \bibinfo{journal}{Phys. Rev. A} \textbf{\bibinfo{volume}{82}},
  \bibinfo{pages}{043629} (\bibinfo{year}{2010}).

\bibitem[{\citenamefont{Mahmud and Tiesinga}(2013)}]{Mahmud:2013wf}
\bibinfo{author}{\bibfnamefont{K.~W.} \bibnamefont{Mahmud}} \bibnamefont{and}
  \bibinfo{author}{\bibfnamefont{E.}~\bibnamefont{Tiesinga}},
  \bibinfo{journal}{arXiv:1304.7565}  (\bibinfo{year}{2013}).

\bibitem[]{Jordan:2008}
\bibinfo{author}{\bibfnamefont{S.} \bibnamefont{Jordan}}
  \bibnamefont{and} \bibinfo{author}{\bibfnamefont{E.}
  \bibnamefont{Farhi}}, \bibinfo{journal}{Phys. Rev. A}
  \textbf{\bibinfo{volume}{77}}, \bibinfo{pages}{062329} (\bibinfo{year}{2008}).
  
  
\bibitem[{\citenamefont{Schoelkopf and Girvin}(2008)}]{Schoelkopf:2008p8712}
\bibinfo{author}{\bibfnamefont{R.~J.} \bibnamefont{Schoelkopf}}
  \bibnamefont{and} \bibinfo{author}{\bibfnamefont{S.~M.}
  \bibnamefont{Girvin}}, \bibinfo{journal}{Nature}
  \textbf{\bibinfo{volume}{451}}, \bibinfo{pages}{664} (\bibinfo{year}{2008}).

\bibitem[{\citenamefont{Devoret and Schoelkopf}(2013)}]{Stajic:2013dh}
\bibinfo{author}{\bibfnamefont{M.~H.} \bibnamefont{Devoret}} \bibnamefont{and}
  \bibinfo{author}{\bibfnamefont{R.~J.} \bibnamefont{Schoelkopf}},
  \bibinfo{journal}{Science} \textbf{\bibinfo{volume}{339}},
  \bibinfo{pages}{1169} (\bibinfo{year}{2013}).

\bibitem[{\citenamefont{Manucharyan et~al.}(2009)\citenamefont{Manucharyan,
  Koch, Glazman, and Devoret}}]{Manucharyan:2009fo}
\bibinfo{author}{\bibfnamefont{V.~E.} \bibnamefont{Manucharyan}},
  \bibinfo{author}{\bibfnamefont{J.}~\bibnamefont{Koch}},
  \bibinfo{author}{\bibfnamefont{L.~I.} \bibnamefont{Glazman}},
  \bibnamefont{and} \bibinfo{author}{\bibfnamefont{M.~H.}
  \bibnamefont{Devoret}}, \bibinfo{journal}{Science}
  \textbf{\bibinfo{volume}{326}}, \bibinfo{pages}{113} (\bibinfo{year}{2009}).

\bibitem[{\citenamefont{Koch et~al.}(2007)\citenamefont{Koch, Yu, Gambetta,
  Houck, Schuster, Majer, Blais, Devoret, Girvin, and
  Schoelkopf}}]{Transom:2007}
\bibinfo{author}{\bibfnamefont{J.}~\bibnamefont{Koch}},
  \bibinfo{author}{\bibfnamefont{T.}~\bibnamefont{Yu}},
  \bibinfo{author}{\bibfnamefont{J.}~\bibnamefont{Gambetta}},
  \bibinfo{author}{\bibfnamefont{A.}~\bibnamefont{Houck}},
  \bibinfo{author}{\bibfnamefont{D.}~\bibnamefont{Schuster}},
  \bibinfo{author}{\bibfnamefont{J.}~\bibnamefont{Majer}},
  \bibinfo{author}{\bibfnamefont{A.}~\bibnamefont{Blais}},
  \bibinfo{author}{\bibfnamefont{M.}~\bibnamefont{Devoret}},
  \bibinfo{author}{\bibfnamefont{S.~M.} \bibnamefont{Girvin}},
  \bibnamefont{and}
  \bibinfo{author}{\bibfnamefont{R.}~\bibnamefont{Schoelkopf}},
  \bibinfo{journal}{Phys. Rev. A} \textbf{\bibinfo{volume}{76}},
  \bibinfo{pages}{42319} (\bibinfo{year}{2007}).

\bibitem[{\citenamefont{Zakka-Bajjani et~al.}(2011)\citenamefont{Zakka-Bajjani,
  Nguyen, Lee, Vale, Simmonds, and Aumentado}}]{ZakkaBajjani:2011in}
\bibinfo{author}{\bibfnamefont{E.}~\bibnamefont{Zakka-Bajjani}},
  \bibinfo{author}{\bibfnamefont{F.}~\bibnamefont{Nguyen}},
  \bibinfo{author}{\bibfnamefont{M.}~\bibnamefont{Lee}},
  \bibinfo{author}{\bibfnamefont{L.~R.} \bibnamefont{Vale}},
  \bibinfo{author}{\bibfnamefont{R.~W.} \bibnamefont{Simmonds}},
  \bibnamefont{and}
  \bibinfo{author}{\bibfnamefont{J.}~\bibnamefont{Aumentado}},
  \bibinfo{journal}{Nat. Phys.} \textbf{\bibinfo{volume}{7}},
  \bibinfo{pages}{599} (\bibinfo{year}{2011}).

\bibitem[{\citenamefont{Koch et~al.}(2010)\citenamefont{Koch, Houck, Hur, and
  Girvin}}]{Koch:2010}
\bibinfo{author}{\bibfnamefont{J.}~\bibnamefont{Koch}},
  \bibinfo{author}{\bibfnamefont{A.~A.} \bibnamefont{Houck}},
  \bibinfo{author}{\bibfnamefont{K.~L.} \bibnamefont{Hur}}, \bibnamefont{and}
  \bibinfo{author}{\bibfnamefont{S.~M.} \bibnamefont{Girvin}},
  \bibinfo{journal}{Phys. Rev. A} \textbf{\bibinfo{volume}{82}},
  \bibinfo{pages}{043811} (\bibinfo{year}{2010}).

\bibitem[{\citenamefont{Kapit}(2013)}]{Kapit:2013wx}
\bibinfo{author}{\bibfnamefont{E.}~\bibnamefont{Kapit}},
  \bibinfo{journal}{arXiv:1302.6596}  (\bibinfo{year}{2013}).

\bibitem[{\citenamefont{Hafezi et~al.}(2011)\citenamefont{Hafezi, Demler,
  Lukin, and Taylor}}]{Hafezi:2011delay}
\bibinfo{author}{\bibfnamefont{M.}~\bibnamefont{Hafezi}},
  \bibinfo{author}{\bibfnamefont{E.~A.} \bibnamefont{Demler}},
  \bibinfo{author}{\bibfnamefont{M.~D.} \bibnamefont{Lukin}}, \bibnamefont{and}
  \bibinfo{author}{\bibfnamefont{J.~M.} \bibnamefont{Taylor}},
  \bibinfo{journal}{Nat. Phys.} \textbf{\bibinfo{volume}{7}},
  \bibinfo{pages}{907} (\bibinfo{year}{2011}).

\bibitem[{\citenamefont{Umucalilar and Carusotto}(2011)}]{Umucalilar:2011}
\bibinfo{author}{\bibfnamefont{R.~O.} \bibnamefont{Umucalilar}}
  \bibnamefont{and}
  \bibinfo{author}{\bibfnamefont{I.}~\bibnamefont{Carusotto}},
  \bibinfo{journal}{Phys. Rev. A} \textbf{\bibinfo{volume}{84}},
  \bibinfo{pages}{043804} (\bibinfo{year}{2011}).
  
  
\bibitem[]{Laughlin:1989kd}
\bibinfo{author}{\bibfnamefont{R.~B.} \bibnamefont{Lauhglin}},
  \bibinfo{journal}{Ann. Phys. (Leipzig)} \textbf{\bibinfo{volume}{191}},
  \bibinfo{pages}{163} (\bibinfo{year}{1989}).  
  

\bibitem[{\citenamefont{Hafezi et~al.}(2007)\citenamefont{Hafezi, Sorensen,
  Demler, and Lukin}}]{Hafezi:PRA2007}
\bibinfo{author}{\bibfnamefont{M.}~\bibnamefont{Hafezi}},
  \bibinfo{author}{\bibfnamefont{A.~S.} \bibnamefont{Sorensen}},
  \bibinfo{author}{\bibfnamefont{E.}~\bibnamefont{Demler}}, \bibnamefont{and}
  \bibinfo{author}{\bibfnamefont{M.~D.} \bibnamefont{Lukin}},
  \bibinfo{journal}{Phys. Rev. A} \textbf{\bibinfo{volume}{76}},
  \bibinfo{pages}{023613} (\bibinfo{year}{2007}).

\bibitem[{\citenamefont{Kapit and Mueller}(2010)}]{Kapit:2010p47218}
\bibinfo{author}{\bibfnamefont{E.}~\bibnamefont{Kapit}} \bibnamefont{and}
  \bibinfo{author}{\bibfnamefont{E.}~\bibnamefont{Mueller}},
  \bibinfo{journal}{Phys. Rev. Lett.} \textbf{\bibinfo{volume}{105}},
  \bibinfo{pages}{215303} (\bibinfo{year}{2010}).
  
\bibitem[{\citenamefont{Kapit et~al.}(2012)\citenamefont{Kapit, Ginsparg, and
  Mueller}}]{Kapit:2012bn}
\bibinfo{author}{\bibfnamefont{E.}~\bibnamefont{Kapit}},
  \bibinfo{author}{\bibfnamefont{P.}~\bibnamefont{Ginsparg}}, \bibnamefont{and}
  \bibinfo{author}{\bibfnamefont{E.}~\bibnamefont{Mueller}},
  \bibinfo{journal}{Phys. Rev. Lett.} \textbf{\bibinfo{volume}{108}},
  \bibinfo{pages}{066802} (\bibinfo{year}{2012}).

\bibitem[{\citenamefont{Umucalilar and
  Carusotto}(2012{\natexlab{a}})}]{Umucalilar:2011b}
\bibinfo{author}{\bibfnamefont{R.~O.} \bibnamefont{Umucalilar}}
  \bibnamefont{and}
  \bibinfo{author}{\bibfnamefont{I.}~\bibnamefont{Carusotto}},
  \bibinfo{journal}{Phys. Rev. Lett.} \textbf{\bibinfo{volume}{108}},
  \bibinfo{pages}{206809} (\bibinfo{year}{2012}{\natexlab{a}}).

\bibitem[{\citenamefont{Hayward et~al.}(2012)\citenamefont{Hayward, Martin, and
  Greentree}}]{Hayward:2012}
\bibinfo{author}{\bibfnamefont{A.~L.~C.} \bibnamefont{Hayward}},
  \bibinfo{author}{\bibfnamefont{A.~M.} \bibnamefont{Martin}},
  \bibnamefont{and} \bibinfo{author}{\bibfnamefont{A.~D.}
  \bibnamefont{Greentree}}, \bibinfo{journal}{Phys. Rev. Lett.}
  \textbf{\bibinfo{volume}{108}}, \bibinfo{pages}{223602}
  (\bibinfo{year}{2012}).

\bibitem[{\citenamefont{Hafezi et~al.}(2013)\citenamefont{Hafezi, Lukin, and
  Taylor}}]{Hafezi:2013wh}
\bibinfo{author}{\bibfnamefont{M.}~\bibnamefont{Hafezi}},
  \bibinfo{author}{\bibfnamefont{M.~D.} \bibnamefont{Lukin}}, \bibnamefont{and}
  \bibinfo{author}{\bibfnamefont{J.~M.} \bibnamefont{Taylor}},
  \bibinfo{journal}{New Journal of Physics} \textbf{\bibinfo{volume}{15}},
  \bibinfo{pages}{063001} (\bibinfo{year}{2013}).

\bibitem[{\citenamefont{Shanks et~al.}(2013)\citenamefont{Shanks, Underwood,
  and Houck}}]{Shanks:2013hi}
\bibinfo{author}{\bibfnamefont{W.~E.} \bibnamefont{Shanks}},
  \bibinfo{author}{\bibfnamefont{D.~L.} \bibnamefont{Underwood}},
  \bibnamefont{and} \bibinfo{author}{\bibfnamefont{A.~A.} \bibnamefont{Houck}},
  \bibinfo{journal}{Nature Communications} \textbf{\bibinfo{volume}{4}},
  (\bibinfo{year}{2013}).

\bibitem[{\citenamefont{Carusotto et~al.}(2009)\citenamefont{Carusotto, Gerace,
  Tureci, De~Liberato, Ciuti, and Imamo{\v g}lu}}]{Carusotto:2009}
\bibinfo{author}{\bibfnamefont{I.}~\bibnamefont{Carusotto}},
  \bibinfo{author}{\bibfnamefont{D.}~\bibnamefont{Gerace}},
  \bibinfo{author}{\bibfnamefont{H.}~\bibnamefont{Tureci}},
  \bibinfo{author}{\bibfnamefont{S.}~\bibnamefont{De~Liberato}},
  \bibinfo{author}{\bibfnamefont{C.}~\bibnamefont{Ciuti}}, \bibnamefont{and}
  \bibinfo{author}{\bibfnamefont{A.}~\bibnamefont{Imamo{\v g}lu}},
  \bibinfo{journal}{Phys. Rev. Lett.} \textbf{\bibinfo{volume}{103}},
  \bibinfo{pages}{033601} (\bibinfo{year}{2009}).

\bibitem[{\citenamefont{Hafezi et~al.}(2012)\citenamefont{Hafezi, Chang,
  Gritsev, Demler, and Lukin}}]{Hafezi:2012ku}
\bibinfo{author}{\bibfnamefont{M.}~\bibnamefont{Hafezi}},
  \bibinfo{author}{\bibfnamefont{D.}~\bibnamefont{Chang}},
  \bibinfo{author}{\bibfnamefont{V.}~\bibnamefont{Gritsev}},
  \bibinfo{author}{\bibfnamefont{E.}~\bibnamefont{Demler}}, \bibnamefont{and}
  \bibinfo{author}{\bibfnamefont{M.}~\bibnamefont{Lukin}},
  \bibinfo{journal}{Phys. Rev. A} \textbf{\bibinfo{volume}{85}},
    \bibinfo{pages}{ 013822}(\bibinfo{year}{2012}).
  
 

\bibitem[{\citenamefont{Nunnenkamp et~al.}(2011)\citenamefont{Nunnenkamp, Koch,
  and Girvin}}]{Nunnenkamp:2011dm}
\bibinfo{author}{\bibfnamefont{A.}~\bibnamefont{Nunnenkamp}},
  \bibinfo{author}{\bibfnamefont{J.}~\bibnamefont{Koch}}, \bibnamefont{and}
  \bibinfo{author}{\bibfnamefont{S.~M.} \bibnamefont{Girvin}},
  \bibinfo{journal}{New Journal of Physics} \textbf{\bibinfo{volume}{13}},
  \bibinfo{pages}{095008} (\bibinfo{year}{2011}).

\bibitem[{\citenamefont{Schir{\'o} et~al.}(2012)\citenamefont{Schir{\'o},
  Bordyuh, {\"O}ztop, and Tureci}}]{schiro:2012}
\bibinfo{author}{\bibfnamefont{M.}~\bibnamefont{Schir{\'o}}},
  \bibinfo{author}{\bibfnamefont{M.}~\bibnamefont{Bordyuh}},
  \bibinfo{author}{\bibfnamefont{B.}~\bibnamefont{{\"O}ztop}},
  \bibnamefont{and} \bibinfo{author}{\bibfnamefont{H.}~\bibnamefont{Tureci}},
  \bibinfo{journal}{Phys. Rev. Lett.} \textbf{\bibinfo{volume}{109}},
  \bibinfo{pages}{053601} (\bibinfo{year}{2012}).

\bibitem[{\citenamefont{Umucalilar and
  Carusotto}(2012{\natexlab{b}})}]{Umucalilar:2012vw}
\bibinfo{author}{\bibfnamefont{R.~O.} \bibnamefont{Umucalilar}}
  \bibnamefont{and}
  \bibinfo{author}{\bibfnamefont{I.}~\bibnamefont{Carusotto}},
  \bibinfo{journal}{arXiv:1210.3070}  (\bibinfo{year}{2012}{\natexlab{b}}).


\end{thebibliography}

\end{document}